\documentclass[journal=jacsat,manuscript=article]{achemso}

\usepackage{chemformula} 
\usepackage[T1]{fontenc} 
\usepackage{booktabs}  

\author{Philippe Schwaller}
\affiliation[IBM Research]
{IBM Research GmbH, Zurich, Switzerland}
\alsoaffiliation[University of Cambridge]
{Department of Physics, University of Cambridge, Cambridge, United Kingdom}
\email{phs@zurich.ibm.ch}
\author{Teodoro Laino}
\affiliation[IBM Research]
{IBM Research GmbH, Zurich, Switzerland}
\author{Th\'eophile Gaudin}
\affiliation[IBM Research]
{IBM Research GmbH, Zurich, Switzerland}
\author{Peter Bolgar}
\affiliation[University of Cambridge]
{Department of Chemistry, University of Cambridge, Cambridge, United Kingdom}
\author{Costas Bekas}
\affiliation[IBM Research]
{IBM Research GmbH, Zurich, Switzerland}
\author{Alpha A Lee}
\affiliation[University of Cambridge]
{Department of Physics, University of Cambridge, Cambridge, United Kingdom}
\email{aal44@cam.ac.uk}

\title[An \textsf{achemso} demo]
  {Molecular Transformer -- A Model for Uncertainty-Calibrated Chemical Reaction Prediction}

\abbreviations{IR,NMR,UV}

\begin{document}

\begin{abstract}
Organic synthesis is one of the key stumbling blocks in medicinal chemistry. A necessary yet unsolved step in planning synthesis is solving the forward problem: given reactants and reagents, predict the products. Similar to other work, we treat reaction prediction as a machine translation problem between SMILES strings of reactants-reagents and the products. We show that a multi-head attention Molecular Transformer model outperforms all algorithms in the literature, achieving a top-1 accuracy above 90\% on a common benchmark dataset. Our algorithm requires no handcrafted rules, and accurately predicts subtle chemical transformations. Crucially, our model can accurately estimate its own uncertainty, with an uncertainty score that is 89\% accurate in terms of classifying whether a prediction is correct. Furthermore, we show that the model is able to handle inputs without reactant-reagent split and including stereochemistry, which makes our method universally applicable. 
\end{abstract}


\section{Introduction} 

Organic synthesis -- the making of complex molecules from simpler building blocks -- remains one of the key stumbling blocks in drug discovery \cite{blakemore2018organic}. Although the number of reported molecules has reached 135 million, this still represents only a small proportion of the estimated $10^{60}$ feasible drug-like compounds \cite{bohacek1996art, bostrom2018expanding}. The lack of a synthetic route hinders access to potentially fruitful regions of chemical space. Tackling the challenge of organic synthesis with data-driven approaches is particularly timely as generative models in machine learning for molecules are coming of age \cite{kusner2017grammar, gomez2018automatic, griffiths2017constrained,popova2018deep, blaschke2018application,jin2018junction,kang2018conditional}. These generative models enrich the toolbox of medicinal chemistry by suggesting potentially promising molecules that lie outside of known scaffolds.

There are three salient challenges in predicting chemical reactivity and designing organic synthesis. First, simple combinatorics would suggest that the space of possible reactions is even greater than the already intractable space of possible molecules. As such, strategies that involve handcrafted rules quickly become intractable. Second, reactants seldom contain only one reactive functional group. Designing synthesis requires one to predict which functional group will react with a particular reactant and where a reactant will react within a functional group.  Predicting those subtle reactivity differences is challenging because they are often dependent on the what other functional groups are nearby. In addition, for chiral organic molecules, predicting the relative and absolute configuration of chiral centers adds another layer of complexity. Third, organic synthesis is almost always a multistep process where one failed step could invalidate the entire synthesis. For example, the pioneering total synthesis of the antibiotic tetracycline takes 18 steps \cite{charest2005convergent}; even a hypothetical method that would be correct 80\% of the time would only have a 1\% chance of getting 18 predictions correct in a row (assuming independence). Therefore, tackling the synthesis challenge requires methods that are both accurate and have a good uncertainty estimates. This would crucially allow us to estimate the ``risk'' of the proposed synthesis path, and put the riskier steps in the beginning of the synthesis, so that one can fail fast and fail cheap. 

The long history of computational chemical reaction prediction has been extensively reviewed in \cite{engkvist2018computational} and \cite{coley2018machine}. Methods in the literature may be divided into two different groups, namely, template-based and template-free. 

Template-based methods \cite{wei2016neural, coley2017prediction, segler2017neural} use a library of reaction templates or rules. These templates describe the atoms and their bonds in the neighborhood of the reaction center, before and after the chemical reaction has occurred. Template-based methods then consider all possible reactions centers in a molecule, and enumerate the possible transformations based on the templates together with how likely each transformation is to take place. As such, the key steps in all template-based methods are the construction of templates, and the evaluation of how likely the template is to apply. The focus of the literature has thus far been on the latter question of predicting whether a template applies \cite{coley2017prediction,segler2017neural}. However, the problem with the template-based paradigm is that templates themselves are often of questionable validity. Earlier methods generated templates by hand using chemical intuition
\cite{corey1985computer, szymkuc2016computer, grzybowski2018chematica}. Handcrafting is obviously not scalable as the number of reported organic reactions constantly increases and significant time investment is needed to keep up with the literature. Recent machine learning approaches employ template libraries that are automatically extracted from datasets of reactions \cite{coley2017prediction,segler2017neural}. Unfortunately, automatic template extraction algorithms still suffer from having to rely on meta-heuristics to define different ``classes'' of reactions. More problematically, all automatic template extraction algorithms rely on pre-existing atom mapping -- a scheme that maps atoms in the reactants to atoms in the product. However, correctly mapping the product back to the reactant atoms is still an unsolved problem \cite{chen2013automatic} and, more disconcertingly, commonly used tools to find the atom-mapping (e.g. NameRXN \cite{NameRXN, schneider2015development}) are themselves based on libraries of expert rules and templates. This creates a vicious circle -- atom-mapping is based on templates and templates are based on atom-mapping, and ultimately, seemingly automatic techniques are actually premised on handcrafted and often artisanal chemical rules. 

To overcome the limitations of template-based approaches, several template-free methods have emerged over the recent years. Those methods can in turn be categorized into graph-based and sequence-based. Jin et al. characterize chemical reactions by graph edits that lead from the reactants to the products \cite{jin2017predicting}. Their reaction prediction is a two-step process. The first network takes a graph representation of the reactants as input and predicts reactivity scores. Based on those reactivity scores, product candidates are generated and then ranked by a second network. An improved version, where candidates with up to 5 bond changes are taken into account and multi-dimensional reactivity matrices are generated, was recently presented \cite{coley2018}. While an earlier version of the model included both reactants and reagents in the reaction center determination step, the accuracy was significantly improved by excluding the reagents from the reactivity score prediction in the more recent versions. This requires the user to know what are the identity of the reagents, which implicitly means that the user must already know the product as the reagent is defined as chemical species that do not appear in the product! Similarly, Bradshaw et al. \cite{bradshaw2018predicting} separated reactants and reagents and included the reagents only in a context vector for their gated graph neural network. They represented the reaction prediction problem as a stepwise rearrangement of electrons in the reactant molecules. A side effect of phrasing reaction prediction as predicting electron flow is that a preprocessing step must be applied to eliminate reactions where the electron flow cannot easily  be identified -- Bradshaw et al. considered only a subset of the USPTO\_MIT dataset, containing only 73\% of the reactions with a linear electron flow (LEF) topology, thus by definition excluding pericyclic reactions and other important workhorse organic reactions. A more general version of the algorithm was recently presented in \cite{anonymous2019graph}. Perhaps most intriguingly, all graph-based template-free methods in the literature require atom-mapped datasets to generate the ground truth for training, and atom mapping algorithms make use of reaction templates. 

Sequence-based techniques have emerged as an alternative to graph-based methods. The key idea is to use a text representation of the reactants, reagents and products (usually SMILES), and treat reaction prediction as machine translation from one language (reactants--reagents) to another language (products). The idea of applying sequence-based models to the reaction prediction problem was first explored by Nam \& Kim \cite{nam2016linking}. Schwaller et al. \cite{schwaller2018found} have shown that using analogies between organic chemistry and human languages sequence-to-sequence models (seq-2-seq) could compete against graph-based methods. Both previous seq-2-seq works were based on recurrent neural networks for the encoder and the decoder, with one single-head attention layer in-between \cite{bahdanau2014neural, luong2015effective}. Moreover, both previous seq-2-seq forward prediction works separated reactants and reagents in the inputs using the atom-mapping, and \cite{schwaller2018found} tokenized the reagent molecules as individual tokens. To increase the interpretability of the model, Schwaller et al. \cite{schwaller2018found} used attention weight matrices and confidence scores that were generated together with the most likely product. 

In this work, we focus on the question of predicting products given reactants and reagent. We show that a fully attention-based model adapted from \cite{vaswani2017attention} with the SMILES \cite{weininger1988smiles, weininger1989smiles} representation, the Molecular Transformer, outperforms all previous methods while being completely atom-mapping independent and not requiring splitting the input into reactants and reagents. Our algorithm  reaches 90.4\% top-1 accuracy (93.7\% top-2 accuracy) on a common benchmark dataset. Importantly, our algorithm does not make use of any handcrafted rules. It can accurately predict subtle and selective chemical transformations, getting the correct chemoselectivity, regioselectivity and, to some extent, stereoselectivity. In addition, our model can estimate its own uncertainty. The uncertainty score predicted by the model has an ROC-AUC of 0.89 in terms of classifying whether a reaction is correctly predicted. Our model has been made available since August 2018 in the backend of the IBM RXN for Chemistry \cite{IBMRXN}, a free web-based graphical user interface, and has been used by several thousand organic chemists worldwide to perform more than 40,000 predictions so far.

\section{Data}
Most of the publicly available reaction datasets were derived from the patent mining work of Lowe \cite{lowe2012extraction}, where the chemical reactions were described using a text-based representation called SMILES \cite{weininger1988smiles, weininger1989smiles}. In order to compare to previous work we focus on four datasets. The USPTO\_MIT dataset was filtered and split by Jin et al \cite{jin2017predicting}. This dataset was also used in \cite{schwaller2018found} and adapted to a smaller subset called USPTO\_LEF by Bradshaw et al \cite{bradshaw2018predicting} to make it compatible with their algorithm. In contrast to the MIT and LEF datasets, USPTO\_STEREO \cite{schwaller2018found} underwent less filtering and the stereochemical information was kept. Up to date, only seq-2-seq models were used to predict on USPTO\_STEREO. Stereochemistry adds an additional level of complexity because it requires the models not only to predict molecular graph edge changes, but potentially also changes in node labels. Additionally, we used a non-public time-split test set, extracted from the Pistachio database \cite{Pistachio2017}, to compare the performance on a set containing more diverse reactions against a previous seq-2-seq model \cite{schwaller2018found}. 

\begin{table}[!ht]
\centering
\small
\caption{Dataset splits and preprocessing methods used for the experiments}
    \begin{tabular}{l  rrr  r }
        \toprule  
        Reactions in & train & valid & test & total \\
        \midrule
        USPTO\_MIT set \cite{jin2017predicting} & 409,035 & 30,000 & 40,000& 479,035  \\
        \multicolumn{5}{l}{- No stereochemical information}\\ 
        \midrule
        USPTO\_LEF  \cite{bradshaw2018predicting} & * & * & 29,360 & 349,898 \\
         \multicolumn{5}{l}{- Non-public subset of USPTO\_MIT, without e.g. multi-step reactions}\\ 
         \midrule
        USPTO\_STEREO   \cite{schwaller2018found} & 902,581 & 50,131 & 50,258& 1,002,970 \\
        \multicolumn{5}{l}{- Patent reactions until Sept. 2016, includes stereochemistry}\\ 
        \midrule
        Pistachio\_2017   \cite{schwaller2018found} &  & &15418& 15418 \\
        \multicolumn{5}{l}{- Non-public time split test set, reactions from 2017 taken from Pistachio database \cite{schneider2016big, Pistachio2017} }\\ 
         \toprule
         \multicolumn{5}{c}{Preprocessing methods} \\
        \midrule 
       - separated & \multicolumn{4}{l}{source: COc1c(C)c(C)c(OC)c(C(CCCCC\#CCCO)c2ccccc2)c1C\textbf{>}C.CCO.[Pd]} \\ 
        & \multicolumn{4}{l}{target: COc1c(C)c(C)c(OC)c(C(CCCCCCCCO)c2ccccc2)c1C} \\
        - mixed&  \multicolumn{4}{l}{source: C.CCO.COc1c(C)c(C)c(OC)c(C(CCCCC\#CCCO)c2ccccc2)c1C.[Pd]}\\ 
        & \multicolumn{4}{l}{target: COc1c(C)c(C)c(OC)c(C(CCCCCCCCO)c2ccccc2)c1C} \\
         \bottomrule
    \end{tabular}
    \label{tab:dataset}
\end{table}

Table \ref{tab:dataset} shows an overview of the datasets used in this work and points out the two different preprocessing methods. The \emph{separated} reagents preprocessing means that the reactants (educts), which contribute atoms to the product, are weakly separated by a $>$ token from the reagents (e.g. solvents and catalysts). Reagents take part in the reaction, but do not contribute any atom to the product. So far, in most of the work, the reagents have been separated from the reactants. Jin et al. \cite{jin2017predicting} increased their top-1 accuracy by almost 6\%, when they removed the reagents from the first step, where the reaction centers were predicted. In Schwaller et al.  \cite{schwaller2018found}  the reagents were represented not as individual atoms, but as separate reagent tokens and only included the 76 most common reagents \cite{schneider2016s}. Bradshaw et. al. passed the reagent information as a context vector to their model. In \cite{anonymous2019graph} it was shown that the model performs better when the reagents are tagged as such. Unfortunately, the separation of reactants and reagents is not always obvious. Different tools classify different input molecules as the reactants and hence the reagents will also differ \cite{schneider2016s}. For this reason, we decided to train and test on inputs where the reactants and the reagents were mixed and no distinction was made between the two. We called this method of preprocessing \emph{mixed}. The \emph{mixed} preprocessing makes the reaction prediction task significantly harder, as the model has to determine the reaction center from a larger number of molecules. 

All the reactions used in this work were canonicalized using RDKit \cite{landrum2017rdkit}. The inputs for our model were tokenized with the regular expression found in \cite{schwaller2018found}. In contrast to Schwaller et al. \cite{schwaller2018found}, the reagents were not replaced by reagents tokens, but tokenized in the same way as the reactants. 

\section{The Molecular Transformer}
The model used in this work is based on the transformer architecture \cite{vaswani2017attention}. The model was originally constructed for neural machine translation (NMT) tasks. The main architectural difference compared to seq-2-seq models previously used for reaction prediction \cite{schwaller2018found, nam2016linking}, is that the recurrent neural network component was completely removed and it is fully based on the attention mechanism. 

The transformer is a step-wise autoregressive encoder-decoder model comprised of a combination of multi-head attention layers and positional feed forward layers. In the encoder, the multi-head attention layers attend the input sequence and encode it into a hidden representation. The decoder consists of two types of multi-head attention layers. The first is masked and attends only the preceding outputs of the decoder. The second multi-head attention layer attends encoder outputs, as well as the output of the first decoder attention layer. It basically combines the information of the source sequence with the target sequence that has been produced so far  \cite{vaswani2017attention}.  

A multi-head attention layer itself consists of several scaled-dot attention layers running in parallel, which are then concatenated. The scaled-dot attention layers take three inputs: the keys $K$, the values $V$ and the queries $Q$, and computes the attention as follows:

\begin{equation}
\text{attention}(Q,K,V) = \text{softmax}\left ( \frac{QK^T}{\sqrt{d_k}} \right ) V.
\end{equation}

The dot product of the queries and the keys computes how closely aligned the keys are with the queries. If the query and the key are aligned, their dot product will be large and vice versa. Each key has an associated value vector, which is multiplied with the output of the softmax, through which the dot-products were normalized and the largest components were emphasized. $d_k$ is a scaling factor depending on the layer size. The encoder computes interesting features from the input sequence, which are then queried by the decoder depending on its preceding outputs \cite{vaswani2017attention}. 

One main advantage of the transformer architecture compared to the seq-2-seq models used in \cite{nam2016linking, schwaller2018found} is the multi-head attention, which allows the encoder and decoder to peek at different tokens simultaneously.

Since the recurrent component is missing in the transformer architecture, the sequential nature of the data is encoded with positional encodings \cite{gehring2017convolutional}. Positional encodings add position-dependent trigonometric signals (see Equations \ref{eq_encoding}) to the token embeddings of size $d_{\text{emb}}$ and allow the network to know where the different tokens are situated in the sequence. 

\begin{equation}
\label{eq_encoding}
\text{PE}_{(pos,2i)} = \sin\left(\frac{pos}{10000^{2i/d_{\text{emb}}}} \right), \qquad \text{PE}_{(pos,2i+1)} = \cos \left(\frac{pos}{10000^{2i/d_{\text{emb}}}} \right)
\end{equation}

We based this work on the PyTorch implementation provided by OpenNMT \cite{opennmt}. All the components of the transformer model are explained and illustrated graphically on \cite{Annotated}.

While the base transformer model had 65M parameters \cite{vaswani2017attention}, we decreased the number of trainable weights to 12M by going from 6 layers of size 512 to 4 layers of size 256. We experimented with label smoothing \cite{szegedy2016rethinking} and the number of attention heads. In contrast to the NMT model \cite{vaswani2017attention}, we set the label smoothing parameter to 0.0. As seen below, a non-zero label smoothing parameter encourages the model to be less confident and therefore negatively affects its ability to discriminate between correct and incorrect predictions. Moreover, we observed that at least 4 attention heads were required to achieve peak accuracies. We however, kept the original 8 attention heads, because this configuration achieved superior validation performance. For the training we use the ADAM optimizer \cite{kingma2014adam} and vary the learning rate as described in \cite{vaswani2017attention} using 8000 warm up steps, the batch size is set to approximately 4096 tokens, the gradients are accumulated over four batches and normalized by the number of tokens. The model and results can be found online \cite{github}.

\section{Results \& Discussion}

\begin{table}[h]
  \caption{Ablation study of Molecular Transformer on the USPTO\_MIT dataset with separated reagents. Train and test time were measured on a single Nvidia P100 GPU. The test set contained 40k reactions. }
  \label{tab:trans}
  \centering
  \begin{tabular}{lrrrrrr}
    \toprule
     &Top-1 [\%] & Top-2  [\%]& Top-3  [\%]& Top-5  [\%] & Training & Testing \\
      \midrule
     \multicolumn{7}{c}{Single models}\\ 
     \midrule
Baseline & 88.8  &  92.6  &  93.7  &  94.4  & 24h& 20m \\
Baseline augm. & 89.6 &  93.2 &  94.2 &  95.0 & 24h  & 20m \\
Baseline augm. & 90.1 &  93.5 &  94.4 &  95.2 & 48h & 20m \\
Augm. av. 20 & \textbf{90.4} &  \textbf{93.7} &  \textbf{94.6} &  \textbf{95.3} & 48h & 20m \\
    \midrule
     \multicolumn{7}{c}{Ensemble models}\\ 
     \midrule
Ens. of 5 & 90.5 &  93.8 &  94.8 &  95.5 & 48h & 1h25m \\
Ens. of 10 &90.6 & 93.9 & 94.8 & 95.5 & 48h & 2h40m \\
Ens. of 20 &90.6 & 93.8 & 94.9 & 95.6 & 48h & 5h03m \\
Ens. of 2 av. 20 : & \textbf{91.0} & \textbf{94.3} & \textbf{95.2} & \textbf{95.8} & 2x48h & 32m \\
     
    \bottomrule
  \end{tabular}
\label{ablation}
\end{table}

Table \ref{ablation} shows the performance of the model as a function of different training variations. SMILES data augmentation \cite{bjerrum2017smiles} leads to a significant increase in accuracy. We double the training data by generating a copy of every reaction in the training set, where the molecules were replaced by an equivalent random SMILES (augm.) on the range of datasets and preprocessing methods. Results are also improved by averaging the weights over multiple checkpoints, as suggested in \cite{vaswani2017attention}. Our best single models are obtained by training for 48 hours on one GPU (Nvidia P100), saving one checkpoint every 10,000 time steps and averaging the last 20 checkpoints. Ensembling different models is known to increase the performance of NMT models \cite{liu2018comparable}; however, the performance increase (Ens. of 5 / 10 / 20) is marginal compared to parameter averaging. Nonetheless, ensembling two models which contains the weight average of 20 checkpoints of two independently initialized training runs leads to a top-1 accuracy of 91\%. While a higher accuracy and better uncertainty estimation can be obtained by model ensembles, they come at an additional cost in training and/or test time. The top-5 accuracies of our best single models (weight-average of the 20 last checkpoints) on the different datasets are shown in Table \ref{top-k}. The top-2 accuracy is significantly higher than Top-1, reaching over 93\% accuracy.

\begin{table}[h]
  \caption{The single model top-k accuracy of the Molecular Transformer}
  \label{tab:trans}
  \centering
  \begin{tabular}{llrrrr}
    \toprule
    USPTO*&   & Top-1 [\%] & Top-2  [\%]& Top-3  [\%]& Top-5   [\%]\\
    \midrule
\_MIT & separated & 90.4  & 93.7  & 94.6  & 95.3   \\
\_MIT & mixed & \textbf{88.6}  & \textbf{92.4}  & \textbf{93.5}  & \textbf{94.2}   \\
 \midrule 

\_STEREO & separated & 78.1  & 84.0  & 85.8  & 87.1   \\
\_STEREO & mixed & \textbf{76.2}  & \textbf{82.4}  & \textbf{84.3}  & \textbf{85.8}   \\
    \bottomrule
  \end{tabular}
\label{top-k}
\end{table}

\subsection{Comparison with Previous Work}

As all previous works used single models, we consider only single models trained on the data-augmented versions of the datasets rather than ensembles for the remainder of this paper in order to have a fair comparison. Table \ref{comparison} shows that the Molecular Transformer clearly outperforms all methods in the literature across the different datasets. Crucially, although separating reactant and reagent yields the best model (perhaps unsurprisingly because this separation implies knowledge of the product already), the Molecular Transformer still outperforms the literature when reactant and reagents are mixed. Moreover, our model achieves a reasonable accuracy in the \_STEREO dataset, where stereochemical information is taken into account, whereas  all prior graph-based methods in the literature cannot account for stereochemistry.  

\begin{table}[h]
  \caption{Comparison of Top-1 accuracy in [\%] obtained by the different single model methods on the current benchmark datasets.}
  \label{tab:trans}
  \centering
  \begin{tabular}{llrrrrrr}
    \toprule
    USPTO* &  & S2S & WLDN& ELECTRO & GTPN & WLDN5  & our work \\
   & & \cite{schwaller2018found} &  \cite{jin2017predicting}  & \cite{bradshaw2018predicting} & \cite{anonymous2019graph}  &  \cite{coley2018} &  \\
    \midrule
        \_MIT& separated & 80.3 & 79.6 &  &82.4 & 85.6 &  \textbf{90.4}  \\ \midrule
 \_MIT &    \textbf{mixed} &   &74& &  &&\textbf{88.6}   \\ \midrule \midrule
       \_LEF & separated& &84.0 &87.0 &87.4 & 88.3 & \textbf{92.0} \\ \midrule
        \_LEF & \textbf{mixed} & & & & &  & \textbf{90.3} \\ \midrule\midrule
    \_STEREO & separated &65.4 &&& &&\textbf{78.1}\\ \midrule
    \_STEREO &   \textbf{mixed} &&& &&&\textbf{76.2} \\ \midrule
    \bottomrule
  \end{tabular}
\label{comparison}
\end{table}

\begin{figure}[ht!]
  \centering
    \includegraphics[width=0.6\linewidth]{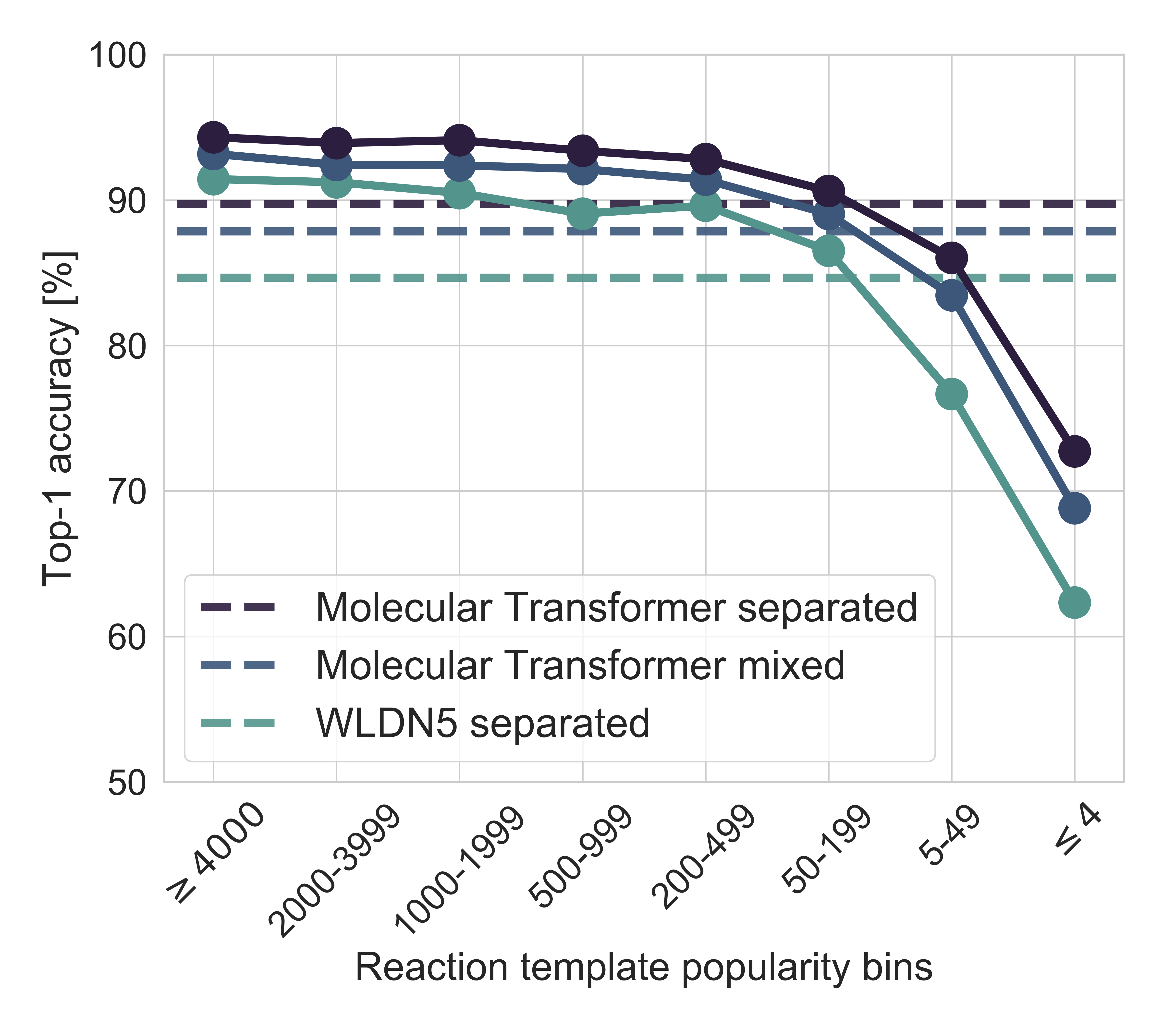}
    \caption{Top-1 accuracy of our augmented mixed and separated USPTO\_MIT single models compared to the model from \cite{coley2018} on the USPTO\_MIT test set, divided into template popularity bins. The dashed lines show the average across all bins.}
 \label{fig:pop}
\end{figure}

Coley et al. \cite{coley2018} published their predictions performance dividing the reactions of the USPTO\_MIT test set into template popularity bins. The template popularity of the test set reactions was computed by counting how many times the corresponding reaction templates were observed in the training set. In Figure \ref{fig:pop} we compare the top-1 accuracy of our USPTO\_MIT models with the model of Coley et al. \cite{coley2018}. Although Coley et al. had separated the reagents in this experiment, we outperform them across all popularity bins even with our model predicting on a mixed reactants-reagents input. The accuracy gap becomes larger as the template popularity decreases. These findings suggest that the Molecular Transformer overfits common reactions less and requires fewer data points to predict well. 

A looming question is how the Molecular Transformer perform by reaction type. Table \ref{Pistachio} shows that the weakest predictions of the Molecular Transformer are on resolutions (the transformation of absolute configuration of chiral centers, where the reagents are often not recorded in the data), and the ominous label of ``unclassified'' (where many mis-transcribed reactions will end up). Moreover, the Molecular Transformer outperforms \cite{schwaller2018found} in virtually every single reaction class. This is because the multi-head attention layer in the  Molecular Transformer can process long ranged interactions between tokens, whereas RNN models impose the inductive bias that tokens fare in sequence space are less related. This bias is erroneous as the token location in SMILES space bears no relation to the distance between atoms in 3D space.

\begin{table}[h]
  \caption{Prediction of the augm. mixed STEREO single model on the Pistachio\_2017 test set, compared to \cite{schwaller2018found}, where the reactants and reagents were separated. }
  \label{tab:trans}
  \centering
  \begin{tabular}{l r r r}
  & Count & S2S acc. \cite{schwaller2018found} [\%]  & Our acc. [\%] \\ \hline
Pistachio\_2017 & 15418 & 60.0 & \textbf{78.0} \\
- Classified & 11817 & 70.2 & \textbf{87.6} \\ \hline
- Heteroatom alkylation and arylation & 2702 & 72.8 & \textbf{86.6} \\
- Acylation and related processes & 2601 & 81.5 & \textbf{90.0} \\
- Deprotections & 1232 & 69.0 & \textbf{88.6} \\
- C-C bond formation & 329 & 55.6 & \textbf{81.2} \\
- Functional group interconversion (FGI) & 315 & 54.0 & \textbf{91.7} \\
- Reductions & 1996 & 71.6 & \textbf{86.1} \\
- Functional group addition (FGA) & 1090 & 71.8 & \textbf{89.3} \\
- Heterocycle formation & 310 & 57.7 & \textbf{90.0} \\
- Protections & 868 & 52.9 & \textbf{87.4} \\
- Oxidations & 339 & 41.3 & \textbf{85.0} \\
- Resolutions & 35 & \textbf{34.3}  & 28.6 \\
- Unrecognized & 3601 & 26.8 & \textbf{46.3} \\ \hline
With stereochemistry & 4103 & 48.2 & \textbf{67.9} \\
Without stereochemistry & 11315 & 64.3 & \textbf{81.6} \\ \hline
Invalid Smiles &  & 2.8 & \textbf{0.5} \\ \hline
  \end{tabular}
  \label{Pistachio}
\end{table}

\subsection{Comparison with human organic chemists}

\begin{figure}[ht!]
  \centering
\includegraphics[width=0.6\linewidth]{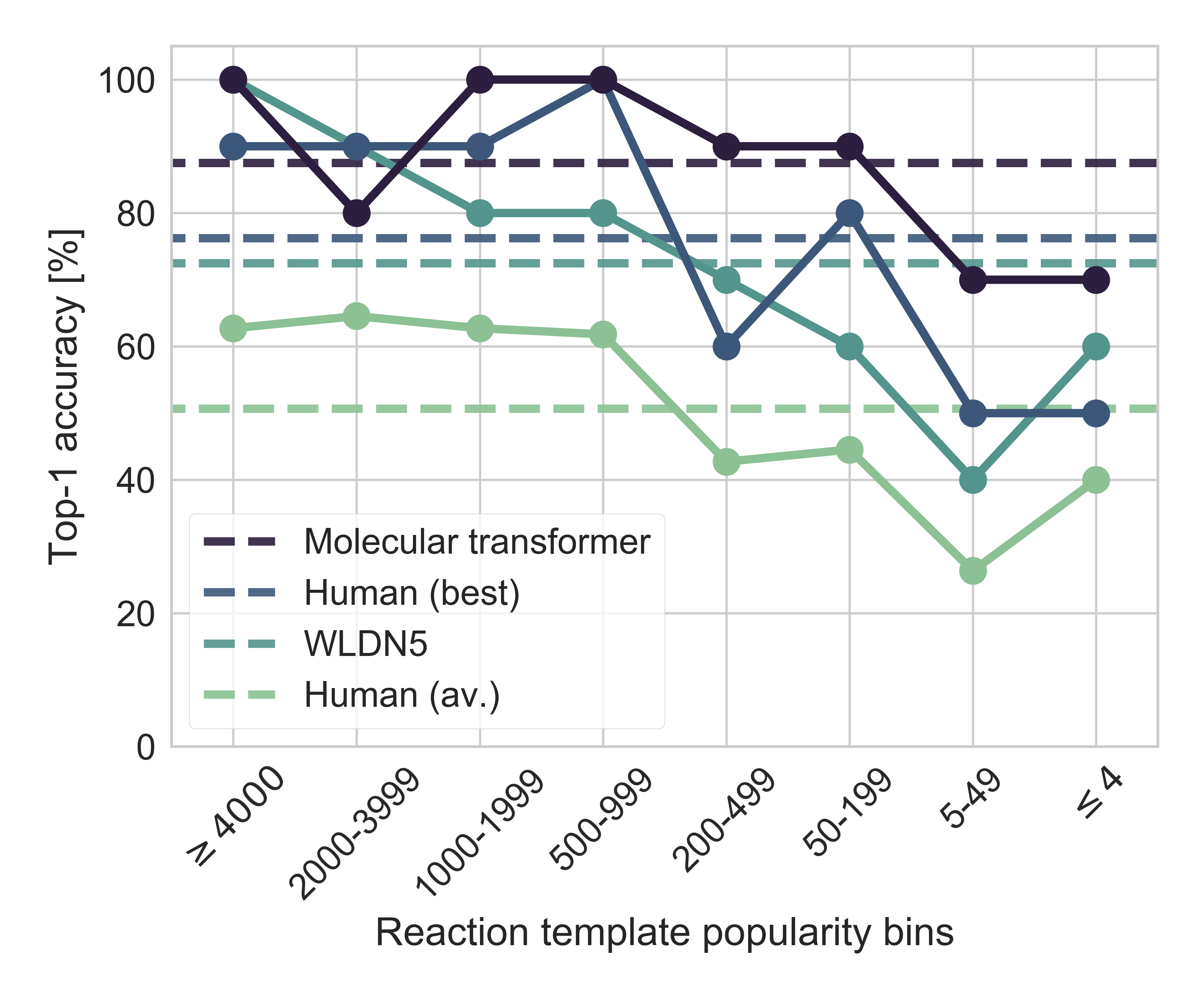}
\centering
\caption{Top-1 accuracy of our model (mixed, USPTO\_MIT) on 80 chemical reactions across 8 reaction popularity bins in comparison with a human study and their graph-based model (WLDN5) \cite{coley2018}.}
\label{fig:human}
\end{figure}

Coley et al. \cite{coley2018} conducted a study, where 80 random reactions from 8 different rarity bins were selected from the USPTO\_MIT test set and presented to 12 chemists (Graduate students to Professors) to predict the most likely outcome. The predictions of the human chemists were then compared against those of the model. We performed the same test with our model trained on the mixed USPTO\_MIT dataset and achieve a top-1 accuracy of 87.5\%, significantly higher than the average of the best human (76.5\%)  and the best graph-based model (72.5\%). Additionally, as seen in Figure \ref{fig:human} Molecular Transformer is generalizable and remains accurate even for the less common reactions. 

\begin{figure}[ht!]
  \centering
  \includegraphics[width=1.0\linewidth]{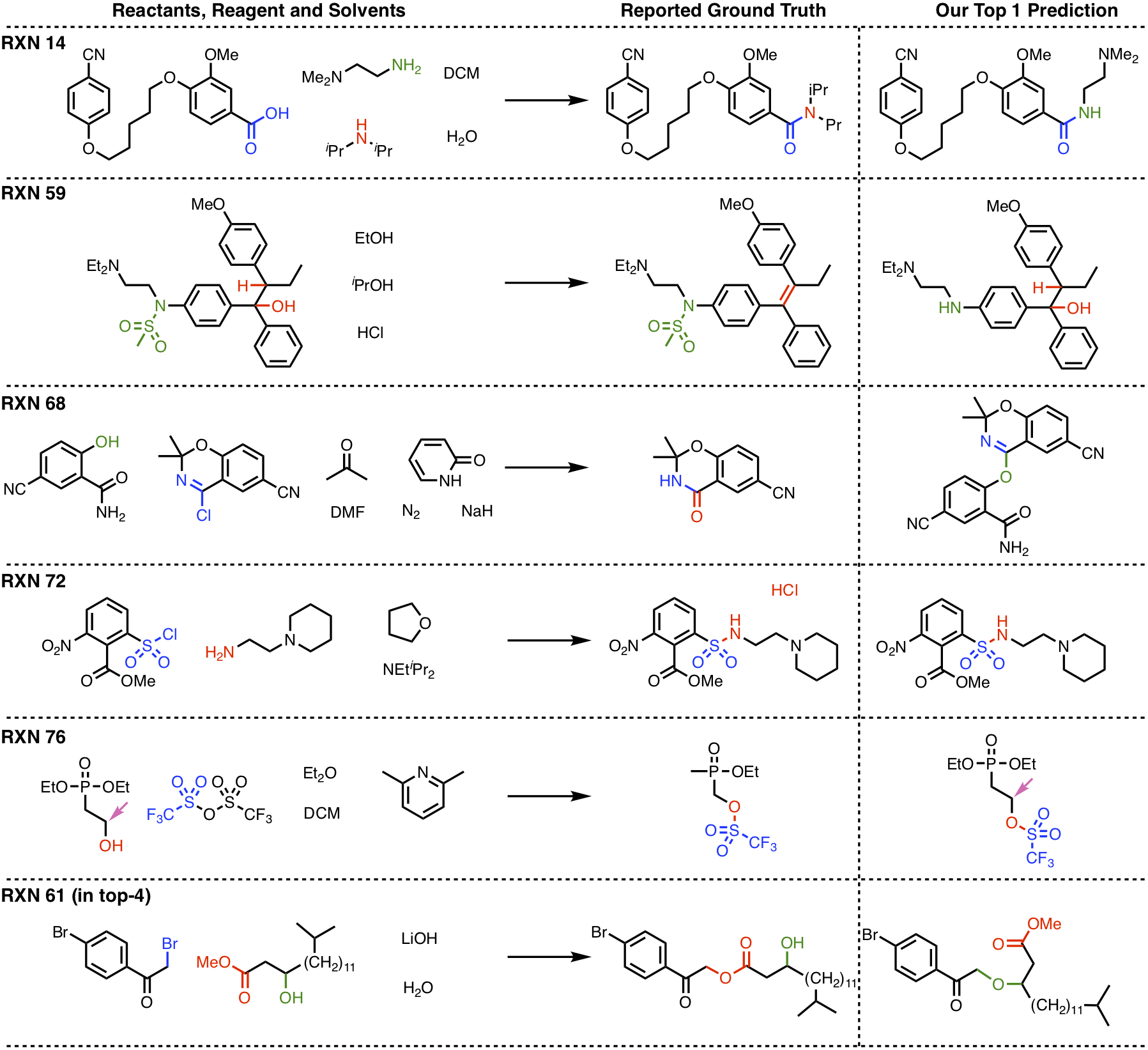}
  \caption{The 6 reactions in the human test set \cite{coley2018} not predicted within top-2 using our model trained on the augmented mixed USPTO\_MIT set. }
  \label{fig:reactions}
\end{figure}

Figure \ref{fig:reactions} shows the 6 of the 80 reactions for which our model did not output the correct prediction in its top-2 choices. Even though our model does not predict the ground truth, it usually predicts a reasonable most likely outcome: In RXN 14, our model predicts that a primary amine acts as the nucleophile in an amide formation reaction rather than a secondary amine, which is reasonable on the grounds of sterics. In RXN 68, the reaction yielding the reported ground truth is via a nucleophilic substitution of \ch{Cl-} by  \ch{OH-} by addition-elimination mechanism, followed by lactim-lactam tautomerism. For the reaction to work there must have been a source of hydroxide ions, which is not indicated among the reactants. In the absence of hydroxide ions, the best nucleophile in the reaction mixture is the phenolate ion generated from the phenol by deprotonation by sodium hydride. In RXN 72, the correct product predicted, but the ground truth additionally reports a by-product (which is mechanistically dubious as \ch{HCl} will react with excess amine to form the ammonium salt). In RXN 76, a carbon atom is clearly missing in the ground truth. In RXN 61, we predict a SN$_2$ where the anion of the alcohol of the beta hydroxy ester acts as nucleophile, whereas the mechanism of the ground truth is presumably ester hydrolysis followed by nucleophilic attack of the carboxylate group. Proton transfers in protic solvents are extremely fast, thus deprotonation of the alcohol \ch{OH} is much faster than ester hydrolysis. Moreover, the carboxylate anion is a poor nucleophile.

\subsection{Uncertainty estimation and reaction pathway scoring}
\begin{figure}[ht!]
  \centering
    \includegraphics[width=0.6\linewidth]{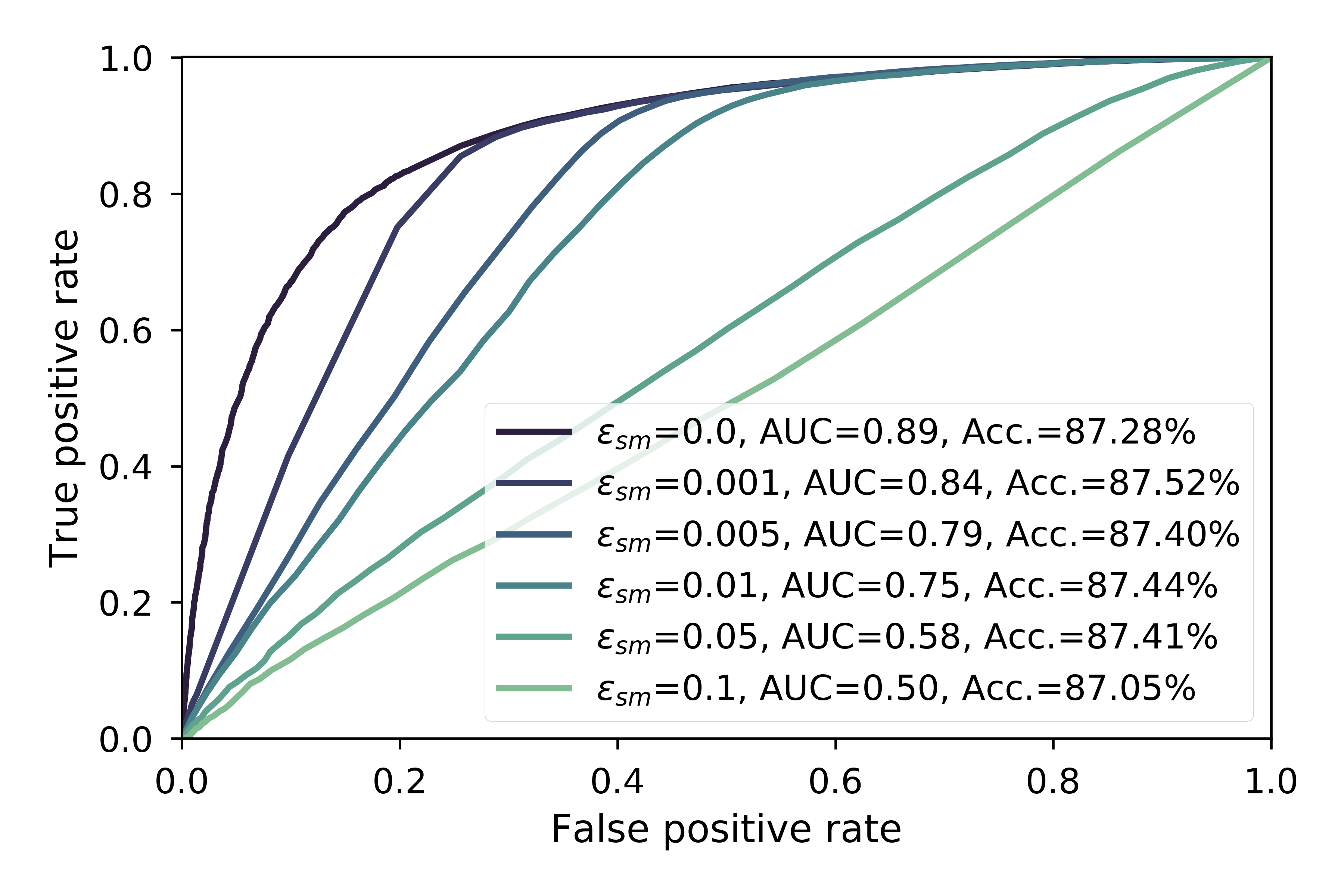}
    \caption{Receiver operatoring characteristic curve for different label smoothing values for a model trained on the mixed USPTO\_MIT dataset, when evaluated on the validation set.}
 \label{fig:smooth}
\end{figure}

As organic synthesis is a multistep process, in order for a reaction predictor to be useful it must be able to estimate its own uncertainty. The Molecular Transformer model provides a natural way achieve this --  the product of the probabilities of all predicted tokens can be used as a confidence score. Figure \ref{fig:smooth} plots the receiver operating characteristics (ROC) curve and shows that the AUC-ROC is 0.89 if we use this confidence score as a threshold to predict whether a reaction is mispredicted. Interestingly, Figure \ref{fig:smooth} reveals that a subtle change in the training method, label smoothing, has a minimal effect on accuracy but a surprisingly significant impact on uncertainty quantification. Label smoothing was introduced by Vaswani et al. \cite{vaswani2017attention} for NMT models. Instead of simply maximizing likelihood of the next target token at a given time step, the network learns a distribution over all possible tokens. Therefore, it is less confident about its predictions. Label smoothing helps to generate higher-scoring translations in terms of accuracy and BLEU score \cite{papineni2002bleu} for human languages, and also helps in terms of reaching higher top-1 accuracy in reaction prediction. The top-1 accuracy on the validation set (mixed, USPTO\_MIT) with the label smoothing parameter set to 0.01 is 87.44\% compared to 87.28\% for no smoothing. However, Figure  \ref{fig:smooth} shows that this small increase in accuracy comes with the cost of not being to able to discriminate between a good and a bad prediction anymore. Therefore, no label smoothing was used during the training of our models. The AUC-ROC of our single mixed USPTO\_MIT model measured on the test set was also at 0.89. The uncertainty estimation metric allows us to estimate the likelihood of a given reactants-product combination, rather only predicting product given reactants, and this could be used as a score to rank reaction pathways \cite{satoh1995sophia, segler2018planning}.

\subsection{Chemically Constrained Beam Search}

As no chemical knowledge was integrated into the model, technically, the model could perform ``alchemy'', e.g. turning a fluoride atom in the reactants into a bromide atom in the products, which was not in the reactants at all. As such, an interesting question is whether the model has learnt to avoid alchemy. To this end, we implemented a constrained beam search, where the probabilities of atomic tokens not observed in the reactants are set to 0.0 and hence not predicted. However, there was no change in accuracy, showing that the model had successfully inferred this constraint from the examples shown during training. 


\section{Conclusion}
We show that a multi-head attention Transformer network, the Molecular Transformer, outperforms all known algorithms in the reaction prediction literature, achieving 90.4\% top-1 accuracy (93.7\% top-2 accuracy) on a common benchmark dataset. The model requires no handcrafted rules, and accurately predicts subtle chemical transformations. Moreover, the Molecular Transformer can also accurately estimate its own uncertainty, with an uncertainty score that is 89\% accurate in terms of classifying whether a prediction is correct. The uncertainty score can be used to rank reaction pathways. We point out that previous work have all considered an unrealistically generous setting of separated reactants and reagents. We demonstrate an accuracy of 88.6\% when no distinction is drawn between reactants and reagents in the inputs, a score that outperforms previous work as well. For the more noisy USPTO\_STEREO dataset, our top-1 accuracies are 78.1\% (separated) and 76.2\% respectively. The Molecular Transformer has been freely available since August 2018 through a graphical user interface on the IBM RXN for Chemistry platform \cite{IBMRXN}, and has so far been used by several thousand organic chemists worldwide for performing more than 40,000 chemical reaction predictions.

\begin{acknowledgement}

PS and AAL acknowledges the Winton Programme for the Physics of Sustainability for funding. The authors thank G. Landrum, R. Sayle, G. Godin and R. Griffiths for useful feedback and discussions. 

\end{acknowledgement}

%
%
%

\bibliography{bib}

\end{document}